\documentclass[12pt]{iopart}

\usepackage{graphicx}
\usepackage{iopams}

\usepackage{subfigure}

\begin{document}

\title[Coherent quantum transport in disordered systems]
       {Coherent quantum transport in disordered systems I:
       The influence of dephasing on the transport properties and absorption
       spectra on one-dimensional systems
       }

\author{
       Jeremy M. Moix, Michael Khasin
\footnote{
       Current address:
       NASA Ames Research Center, Moffett Field, California
}
       and Jianshu Cao
       }
\address{
       Department of Chemistry,
       Massachusetts Institute of Technology,
       77 Massachusetts Avenue, Cambridge, MA 02139
      }\ead{jianshu@mit.edu}

\date{\today}

\begin{abstract}

Excitonic transport in static disordered one dimensional systems is studied
in the presence of thermal fluctuations that are described by the 
Haken-Strobl-Reineker model.
For short times, non-diffusive behavior is observed 
that can be characterized as the free-particle dynamics 
in the Anderson localized system.
Over longer time scales, the environment-induced dephasing is sufficient 
to overcome the Anderson localization caused by the disorder 
and allow for transport to occur which is always seen to be diffusive.
In the limiting regimes of weak and strong dephasing
quantum master equations are developed, and their respective
scaling relations imply the existence of a maximum in the 
diffusion constant as a function of the dephasing rate that
is confirmed numerically.
In the weak dephasing regime, it is demonstrated that the diffusion 
constant is proportional to the square of the localization length 
which leads to a significant enhancement of the transport
rate over the classical prediction. 
Finally, the influence of noise and disorder on the absorption spectrum
is presented and its relationship to the transport properties is discussed.

\end{abstract}

\pacs{71.35.Cc, 71.35.Aa, 72.20.Ee, 72.10.Di, 05.60.Gg}
\maketitle

\section{Introduction}

Understanding and controlling transport in condensed matter systems 
is a topic not only of fundamental scientific interest, but also with 
immediate technological implications.
For example, energy transport is one of the key factors determining 
the efficiency of light harvesting systems, organic photovoltaics,
conducting polymers, and J-aggregate thin 
films~\cite{karl99, podzorov04, sakanoue10, akselrod10, dias09,%
singh09, dykstra08, bednarz03, moix11}.
Energy diffusion has also been shown to play a central role in the rate of 
vibrational relaxation in small molecules and 
DNA~\cite{schofield95, leitner96a, leitner10, goj11}.
Despite this seeming diversity, the underlying physical description of 
these systems is surprisingly similar.
In each case, the transport occurs across a disordered energy landscape that
is modulated by thermal fluctuations from the surrounding environment.
The relative importance of the strength of the disorder and the 
thermal fluctuations to
the electronic coupling may however vary greatly from system to system.
For example, energy transport in J-aggregates are generally characterized 
by relatively weak disorder and exciton-phonon coupling, while
in natural light harvesting systems all of the relevant energy
scales are of comparable magnitude~\cite{bednarz03,cho05}.
This subtle difference in the parameter regimes leads to markedly different
physical behavior in these two systems.

In general the transport properties are relatively well-understood in the
presence of either noise or disorder, but less is known when both are 
present simultaneously.
For example, in the case of static, one- or two-dimensional systems,
Anderson localization occurs for any finite amount of disorder 
leading to a complete lack of transport in the long-time 
limit~\cite{anderson58,ishii73,phillips93,kramer93,ingold04,ping06}.
In the alternative setting of a system subjected to noise but not disorder,
the transport is also --at least qualitatively-- understood.
In particular, the Haken-Strobl-Reineker (HSR) model, 
which describes a quantum system coupled to a high temperature Markovian bath, 
is exactly solvable~\cite{kenkre82,madhukar77,bulatov98,amir09, izrailev97}.
The transport qualitatively behaves like that of a classical Brownian 
process, with an initial period of ballistic expansion before transitioning 
to diffusive transport at long times.

The situation is more complicated when a disordered system
is coupled to an environment.
Thermal excitation from low-lying localized states into more extended
higher energy states can be sufficient to overcome the Anderson localization 
and allow for transport to occur at long times~\cite{thouless81}.
In particular, Mott's theory of variable range hopping
has been quite successful in describing the behavior of the 
low temperature conductivity of disordered systems~\cite{mott69}.
More recently, 
Logan and Wolynes have analyzed the role of dephasing in topologically 
disordered systems and provided approximate scaling relations for the 
transport~\cite{logan87,mukamel89, loring88}.
Their analysis and preliminary numerical results suggest that 
the diffusion coefficient should display a non-monotonic dependence 
on the system-bath coupling strength~\cite{evensky90}.
However, a general description of the transport valid for all temperatures 
is still lacking. 
Alternatively, it has also been shown that an external driving force 
containing a few independent frequency components is equally sufficient to
restore diffusive motion in the long time limit to an otherwise localized
system~\cite{yamada99}.
In summary, while any finite amount of disorder leads
to a lack of diffusion in one or two-dimensional systems, adding
a source of dephasing can be sufficient to allow for transport to occur
by destroying the phase coherence responsible for Anderson localization.

Here, we carry out numerically exact calculations of infinite one-dimensional 
disordered systems over the entire regime of dephasing within the HSR model.
In section~\ref{sec:limits} quantum master equations are derived for the time
evolution of the density matrix in the limiting regimes of weak and strong
dephasing 
These results allow for analytical estimates of the diffusion constant 
and provide an intuitive physical description of the underlying dynamics. 
For large dephasing, it is well known that any coherences created
during the evolution are quickly 
destroyed and the diffusion proceeds by way of classical hopping between
sites~\cite{cao09,hoyer10}.
In the opposite regime of weak dephasing, the exact eigenstates of the system
are accurately approximated by those of the disordered system Hamiltonian,
and coherent quantum transport proceeds via hopping through 
the eigenstates.
This is the regime of phonon-assisted hopping discussed in the early
studies of the conductivity of disordered solids~\cite{miller60}.
These analytical arguments coupled with the exact numerical calculations
lead to several interesting features in the exciton dynamics: 
\begin{itemize}
   \item The strong and weak dephasing master equations allow one to extract
         the respective scaling behavior of the transport properties which
         suggests the existence of a maximum in the 
         diffusion coefficient as a function of the dephasing rate.
         This prediction is confirmed numerically in section~\ref{sec:results}.
         It should be mentioned that similar scaling studies have been 
         carried out in the context of noise-assisted transport in excitonic 
         systems~\cite{lloyd11,wu12b}.
         There, however, the focus is mostly on the optimal energy transfer
         time in finite systems, and the results are largely consistent
         with a purely classical analysis~\cite{wu12}.
         In the preset case there is a genuine and unambiguous quantum 
         aspect to the transport.
   \item In the weak dephasing regime, it is demonstrated that the diffusion
         constant is proportional to the Anderson localization length of the
         disordered system Hamiltonian.
         This implies that for systems weakly coupled to the environment,
         the coherent nature of the transport leads to  
         a diffusion constant that is enhanced by a factor of the localization
         length as compared to the classical prediction.
   \item The third primary result of this work is related to the 
         initial, non-diffusive nature of the transport.  
         Over a timescale on the order of the inverse dephasing rate the
         dynamics are ballistic until the Anderson localization length 
         of the system is reached~\cite{ciuchi11}. 
         The corresponding population probability distribution during this
         period is exponential reflecting the exponential localization of all
         wavefunctions.  
         At longer times, the influence of the dephasing becomes more important
         leading to diffusive motion with a characteristic Gaussian probability
         distribution.
         As a result, the mean-squared displacement can be described
         as a combination of the short-time ballistic transport in the 
         localized system and the long-time diffusive transport.
   \item Finally we present results on the influence of disorder and 
         dephasing on the absorption spectra.
         For many materials, such as photovoltaics, the organic semiconductor 
         must possess not only favorable energy transport characteristics, 
         but also a broad absorption spectrum in order to capture as
         much solar energy as possible.
         It is demonstrated that while reducing the disorder leads to 
         enhanced transport, it also leads to a narrower absorption 
         line shape which is not optimal in this case.
         Therefore, we present a detailed study of the effects of dephasing 
         and disorder on the absorption lineshape.
\end{itemize}

The paper is concluded in section~\ref{sec:applications} where the practical 
implications of these findings are discussed.
For systems such as conducting polymers where the transport
properties are the key factor governing the device performance, then one 
should primarily seek to reduce the amount of static disorder present in the
sample.
Recent experimental results on the conductivity of conducting
polymers and light harvesting systems are discussed in this context.

\section{Model Systems and Scaling Analysis}
\label{Sec:Model}

The HSR model describes quantum transport in the presence
of a classical Markovian environment~\cite{kenkre82}.
At equilibrium all sites are equally populated, and as such, the 
results presented here are strictly valid only in the high temperature limit.
Nevertheless, the model is capable of capturing many of the 
essential physical features of real systems.

In the wavefunction description of the HSR model, the thermal environment
is modeled as a white noise term that modulates the site energies of the 
system Hamiltonian which is described by the stochastic Schr{\"o}dinger 
equation,
\begin{equation}
   i \frac{d}{dt}\left | \psi \right \rangle = H_s \left| \psi\right \rangle
   + \sum_n^N F_n(t) V_n \left | \psi \right \rangle
   \;,
\end{equation}
where $V_n = \left|n\right\rangle\left\langle n \right|$ characterizes
the system-bath coupling in the local basis, and the sum
extends over all $N\rightarrow\infty$ sites of the system. 
The time-dependent factors, $F_n(t)$, are zero-mean Gaussian stochastic
(Wiener) processes with
$\left\langle F_n(t) \right \rangle = 0$, and the autocorrelation,
$\left\langle F_n(t) F_m(t') \right\rangle = \Gamma \delta_{nm}\delta(t-t')$.
In this context, the dephasing rate, $\Gamma$, determines the 
magnitude of the fluctuations.
In order to recover the exact time evolution, the wavefunctions 
must be ensemble averaged over realizations of the noise.

Alternatively, due to the simple noise characteristics of the bath 
the averaging over the environmental fluctuations may be performed 
analytically. 
This procedure leads to an equivalent deterministic 
equation of motion for the density matrix,
\begin{equation}
   \dot \rho(t) = -i\left[ H_s, \rho\right] 
   - \frac{\Gamma}{2}\sum_n \left[V_n, \left[V_n, \rho\right]\,\right]
   \;. \label{eq:HakenStrobl}
\end{equation}
From a numerical standpoint, the density matrix approach is more 
efficient for small systems, while the stochastic implementation becomes 
essential for larger systems.

The bare system is characterized by a one-dimensional tight-binding, 
Anderson Hamiltonian 
\begin{equation}
H_s = \sum_n \epsilon_n c_n^{\dagger} c_n
+ J\left( c_n^\dagger c_{n+1} + c_{n+1}^\dagger c_n\right)
\;,
\end{equation}
where $J$ denotes the electronic coupling strength.
Static disorder is introduced into the system Hamiltonian by taking 
the site energies, $\epsilon_n$, as independent, identically-distributed 
Gaussian random variables characterized by the variance, 
$\sigma^2 = \overline{\left(\epsilon_n - \overline{\epsilon_n}\right)^2}$.
Throughout, the averages over the static disorder are denoted by the 
overline while the quantities denoted by angle brackets characterize the 
quantum-mechanical average over the environment.

The HSR model for the disordered system is completely characterized by 
only three parameters: 
the electronic coupling between sites, $J$, the variance of the 
static disorder, $\sigma^2$, and the dephasing rate, $\Gamma$.
The electronic coupling $J$ is used to set the energy scale throughout, 
which leaves only two independent dimensionless quantities, 
$\Gamma/J$ and $\sigma/J$.
In our simulations, the HSR is always seen to lead to diffusion at long
times and we extract the diffusion constant, $D$, 
from the limiting behavior of the mean-squared displacement (MSD), 
\begin{equation}
   2 D t = \lim_{t\rightarrow\infty}  
   \overline { \left \langle R^2(t) \right \rangle } 
   \;.
\end{equation}
The mean-squared displacement is calculated from,
\begin{equation}
   \left\langle R(t)^2 \right \rangle = \sum_n n^2  \rho_{nn}(t) \;,
\end{equation}
with the origin conveniently chosen such that the first moment is zero.

It is not immediately obvious that the transport should always be
diffusive, particularly in the weak dephasing regime where
the effects of both Anderson localization and quantum coherence are prevalent.
In the classical limit and in the disorder-free case, 
it can be shown analytically that the transport is always diffusive in the 
long time limit.
But outside of these two regimes there is no known 
proof (at least to our knowledge) that the MSD
is guaranteed to increase linearly with time in the steady state. 
However, all of our numerical simulations as well as those of similar earlier
analytical and numerical results suggest that the transport is indeed
diffusive~\cite{thouless81,lee85,evensky90}.

In addition to the transport properties, we also compute the
linear absorption spectrum from the dipole autocorrelation
function~\cite{mukamel99},
\begin{equation}
   A(\omega) \propto {\rm Re}\int_0^\infty dt\, e^{i\omega t} \,
   \overline{\left \langle \mu(t) \mu(0) \right \rangle} \;,
   \label{eq:absorption}
\end{equation}
where the dipole moment operator is given by,
$\mu = \sum_n \mu_n \left( |n\rangle \langle 0| + |0\rangle \langle n|\right)$.
Since all of the molecules in the system are assumed identical, 
we take the magnitude of the individual dipole moments 
equal to the constant value $\mu_n = \mu_0$.

In the numerical simulations presented below the dynamics were performed
in one-dimensional chains of up to $500$ sites and averaged over 
$500$ to $1000$ independent realizations of the static disorder.
The transient short-time effects of the transport properties decay on 
the order of $J/\Gamma$ so that
the dynamics are required for at least an order of magnitude longer
in order to obtain a reliable estimate of the diffusion coefficient.

\subsection{Limiting behavior of the transport properties}
\label{sec:limits}

\subsubsection{Homogeneous systems}

When the disorder is absent ($\sigma=0$), the HSR model is analytically 
solvable.
In this case the MSD is given by~\cite{kenkre82,madhukar77,witkoskie02},
\begin{equation}
   \left\langle R(t)^2 \right \rangle_{\sigma=0} =
   \frac{4 J^2}{\Gamma}
   \left[t - \frac{1}{\Gamma}\left( 1- e^{-\Gamma t}\right) \right] 
   \;, \label{eq:msd}
\end{equation}
which is equivalent to that of a Brownian particle.
For a time of the order of $\Gamma^{-1}$,
the transport displays free-particle, ballistic behavior 
($\left\langle R(t)^2 \right \rangle \propto t^2$)
which transitions to diffusive motion at long times
with the diffusion coefficient,
\begin{equation}
   D_{\rm hom} = 2 J^2/{\Gamma} \;.
   \label{eq:Reineker}
\end{equation}
It is clearly seen that~\eref{eq:Reineker} diverges at small dephasing. 
However, if $\Gamma=0$ exactly, then any amount of disorder 
will lead to Anderson localization and no diffusion.
In this interesting regime of vanishing dephasing but finite disorder
the dynamics are highly coherent and 
display a sensitive dependence on the model parameters.
This region will be explored in depth in the numerical simulations presented
in section~\ref{sec:results}.

\subsubsection{Strong damping regime: hopping rate}

In the presence of both dephasing and disorder, analytic expressions
for the MSD valid for the entire parameter regime are no longer available.
However, if the dephasing rate is sufficiently large such 
that $\Gamma/J \gg 1$ 
then any coherences created during the course of the time evolution 
are quickly destroyed.
In this case,~\eref{eq:HakenStrobl} reduces to a model of classical 
hopping between neighboring sites~\cite{cao09}.
As has been shown previously, averaging the hopping rates over the disorder 
distribution leads to the diffusion 
coefficient~\cite{jayannavar88,cao09,hoyer10},
\begin{equation}
   D_{\rm hop} = \frac{2 J^2 \Gamma}{\Gamma^2 + \sigma^2}
   \;. \label{eq:hopping}
\end{equation}
Notice that~\eref{eq:hopping} displays a non-monotonic dependence 
on $\Gamma$, and in particular, 
optimal transport when $\Gamma_{\rm opt} = \sigma$.
For large dephasing,~\eref{eq:Reineker} is recovered, and 
reduces to $D_{\rm hop} \approx 2 \Gamma J^2/\sigma^2$ at small dephasing.

The linear increase in the diffusion constant for $\Gamma/J \ll 1$ 
has been previously discussed in the context of noise-assisted 
transport~\cite{cao09,hoyer10}.
However,~\eref{eq:hopping} is applicable only when the classical hopping 
description is appropriate.  
In the weak dephasing regime the dynamics are highly coherent so the 
prefactor predicted by the classical hopping description of the dynamics
can not be correct.
As the dephasing decreases, the exact eigenstates of the total system 
rotate from the local basis into the exciton basis, 
and the wavefunctions will delocalize over coherent domains. 
The size of these domains is on the order of the Anderson localization length, 
which, in turn, is governed by the disorder strength~\cite{bednarz03,ingold04}.
The short-time dynamics of an initial localized excitation in the chain 
will be characterized by a rapid expansion filling one of these localized 
domains. 
Over a longer timescale (of the order of $J/\Gamma$)
the dephasing will allow for the collective hopping 
of the delocalized states among the various localization segments.
These qualitative arguments are formalized in the following subsection
through the development of a weak-coupling master equation.

\subsubsection{Weak damping regime: coherent transport}

In the weak dephasing regime, an accurate description of the 
exciton dynamics may be obtained from the Redfield equation.
Since the system is disordered and the diffusion constant depends only 
on the long-time dynamics, the 
secular and Markov approximations can be employed without a significant 
loss of accuracy.
In this case, the exciton populations are governed by the master equation,
\begin{equation}
   \dot \rho_{\kappa\kappa}(t)  = \sum_{\lambda\neq \kappa}  W_{\kappa\lambda} \rho_{\lambda\lambda}(t)
                      - \sum_{\lambda\neq \kappa} W_{\lambda\kappa} \rho_{\kappa\kappa}(t) 
   \;, \label{eq:redfield}
\end{equation}
where the rates,
\begin{equation}
   W_{\kappa\lambda} = W_{\lambda\kappa} = \Gamma \sum_{n} 
            \left| \left \langle \kappa | n \right\rangle
            \left \langle n | \lambda \right \rangle \right|^2
            \;. \label{eq:redfield_rates}
\end{equation}
The Greek labels here denote eigenstates of the static disordered 
system Hamiltonian (exciton states).
The equivalence of the forward and backward rates is a consequence
of the high temperature nature of the thermal environment.
In the secular approximation, the coherences are decoupled from the
populations and their time evolution can be determined analytically as,
$\rho_{\kappa\lambda}(t)= 
e^{-\left( i\omega_{\kappa\lambda} + \Gamma - W_{\kappa \lambda} \right) t}
\rho_{\kappa\lambda}(0)$,
where $\omega_{\kappa\lambda}$ is the difference in eigenenergies 
between states $\kappa$ and $\lambda$.

Equation~\eref{eq:redfield} is a Markovian master equation
that describes hopping through the eigenstates of the disordered system.
The rates --and hence the diffusion constant-- 
increase linearly with $\Gamma$ as with the weak dephasing limit of 
the classical hopping result in~\eref{eq:hopping}.
However, here there is an important difference in that the Redfield rates
are now correctly weighted by the eigenstate overlaps.
The classical hopping rates capture the correct dephasing dependence 
of the diffusion constant, but incorrectly describe the disorder dependence 
of the prefactor.
As will be shown below, this difference is substantial in many
physically relevant situations.

The eigenstate overlaps in the Redfield rates are determined solely by the 
disordered system Hamiltonian. 
Therefore, they depend only on the ratio of $\sigma/J$.
In order to understand this disorder dependence
consider, for example, the rate element $W_{\kappa\lambda}$.
Due to the static disorder, the eigenstate $\lambda$ will be localized 
at a position $n_\lambda$ in the chain with a spatial extent, $\xi_\lambda$, 
and similarly for state $\kappa$.
On average, the spatial extent of $\kappa$ and $\lambda$ is determined by the 
Anderson localization length, $\xi$.
As a result, the rate $W_{\kappa\lambda}$ can be appreciable
only if the states $\kappa$ and $\lambda$ are located at position
in the chain within an Anderson localization length of one another.
That is, $\left| n_\kappa- n_\lambda \right| \lesssim \xi$.
One sees that the fundamental length scale governing the
magnitude of the rates in the Redfield description 
is given by the spatial extent of the wavefunctions, i.e., 
the Anderson localization length.

Hence, on scaling grounds we expect the diffusion constant to 
have the following form,
\begin{equation}
   D_{\rm coh} = \Gamma \xi^2
   \;, \label{eq:redfield_scaling}
\end{equation}
where the factor, $\xi$, is directly proportional to the Anderson 
localization length.
As in the large dephasing limit, the diffusion in the weak dephasing 
regime  may also be interpreted as a hopping process
except that here the fundamental step size is determined by the 
localization length.
The mean-field theory results of Logan and Wolynes have also 
arrived at a similar conclusion~\cite{logan87,evensky90}.
Their approach indicates that the localization length appearing 
in~\eref{eq:redfield_scaling} is, in fact, simply the inverse 
participation ratio.
This result is consistent with our numerical results presented below,
although our preliminary calculations in two-dimensional systems
indicate that this result is only correct up to a scaling constant.

\subsection{Localization Length}\label{sec:ipr}

\begin{figure}
   \begin{center}
   \includegraphics*[width=0.5\textwidth]{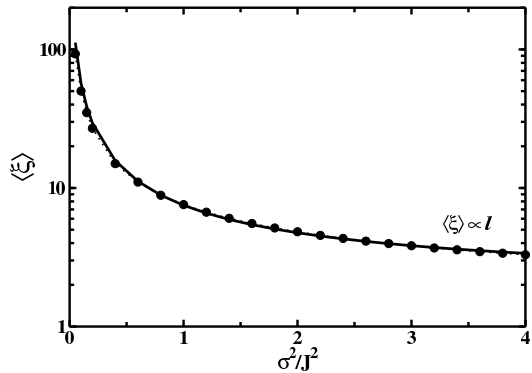}
   \caption{
      The scaling of the inverse participation ratio as a function
      of the disorder strength 
      computed in systems containing up to $4900$ sites.
      The symbols represent the exact numerical results and
      the solid lines depict the fit to the functional form predicted
      from the scaling theory of Anderson localization.
      Specifically, the IPR is fit to the line $\xi\approx 5.5 l $,
      with the mean free path, $l=J^2/\sigma^2$.
   }\label{fig:ipr}
   \end{center}
\end{figure}

Due to the intense interest in the problem of Anderson localization,
accurate scaling relations are known for the localization length
in disordered systems~\cite{phillips93,lee85}.
In one dimension systems, the localization length is directly 
proportional to the mean-free path, $\xi \propto J^2/\sigma^2$. 
In two dimensions the localization length is exponentially larger 
than it is in one dimension, and metallic transport appears at a critical 
value of the disorder strength in three-dimensional systems~\cite{lee85}.
Transport in these systems will be discussing in a future work,
but is beyond the scope of the present paper.
In order to fix the proportionality, we will use the 
inverse participation ratio (IPR) as a proxy for 
the localization length in the disordered chain, 
as suggested by Logan and Wolynes.
The IPR is defined for a particular eigenstate, $\kappa$, as
\begin{equation}
   \xi_\kappa= \left( \sum_n 
                      \left|\left\langle n | \kappa \right\rangle\right|^4
                   \right)^{-1}
   \;.
   \label{eq:ipr}
\end{equation}
Figure~\ref{fig:ipr} presents the scaling of the IPR as a function 
of the disorder computed by direct diagonalization of the disordered system
Hamiltonian.
Due to the high temperature approximation inherent to the HSR model, 
all of the eigenstates are equally capable of contributing to the transport.
Therefore the IPR presented in figure~\ref{fig:ipr}, $\langle \xi \rangle$,
is averaged over all eigenstates,
which reproduce the known functional dependence of the localization length.

\subsection{Implications of the scaling limits}

Based on the limiting behavior of the transport deduced above from
the master equations, several intriguing physical aspects of the 
transport become apparent.
For example, in the weak dephasing regime the quantum coherent nature of the
transport can lead to a considerable enhancement of the diffusion constant
compared with the classical estimate. 
For $\Gamma/J \ll 1$, this ``quantum speedup'' can be determined 
analytically and is given by
\begin{equation}
   D_{\rm coh}/D_{\rm hop} = \frac{\xi^2}{2l} \propto \xi
   \;,\label{eq:speedup}
\end{equation}
which can be quite substantial for reasonable values of the systems 
parameters, as seen from figure~\ref{fig:ipr}.
The enhancement is proportional to the localization
length as a result o fthe enhanced step size of the hopping process 
in the weak dephasing regime.

Secondly, in the weak dephasing regime the transport increases linearly with
$\Gamma$~\eref{eq:redfield} while for strong dephasing the diffusion 
decreases as $\Gamma^{-1}$~\eref{eq:hopping},
which implies that the diffusion constant should be maximal at intermediate 
values of the dephasing.
As mentioned above,~\eref{eq:hopping} also predicts an optimal diffusion
constant at $\Gamma_{\rm opt}=\sigma$. 
However, this estimate is only correct for very large disorder and becomes
increasingly worse as the system becomes more coherent, as will be 
demonstrated in the numerical results presented in the next section.
An improved estimate for the optimal dephasing rate 
may be obtained by equating the proper weak dephasing limit 
of~\eref{eq:redfield_scaling} with the large dephasing result 
in~\eref{eq:Reineker}.
This leads to the improved estimate, 
\begin{equation}
   \Gamma_{\rm max} = \sqrt{2} J/\xi
   \;, \label{eq:D_max}
\end{equation}
which will be shown to be substantially more 
accurate than the prediction provided by~\eref{eq:hopping}.

The existence of a maximum in the diffusion coefficient is a result of
the interplay between the Anderson localization due to static disorder and 
thermal localization originating from the environment~\cite{moix12}.
The optimum occurs in the regime where the exact eigenstates of the total
system transition between the local site basis and the exciton basis.
For example, in an initially Anderson localized system increasing the 
dephasing rate enhances the transport, but it also serves to further localize 
the wavefunctions which reduces the hopping length.
It is this competition that is responsible for the non-monotonic behavior 
of the transport and is generic to many systems.
Similar arguments for the non-monotonic behavior of the transport
have been presented in~\cite{lloyd11} as well as a 
similar estimate for the optimal dephasing rate as~\eref{eq:D_max}.

\section{Numerical Results}\label{sec:results}
\subsection{Diffusion Coefficients}

\begin{figure}
   \begin{center}
   \begin{subfigure}{
         \includegraphics*[width=0.5\textwidth]{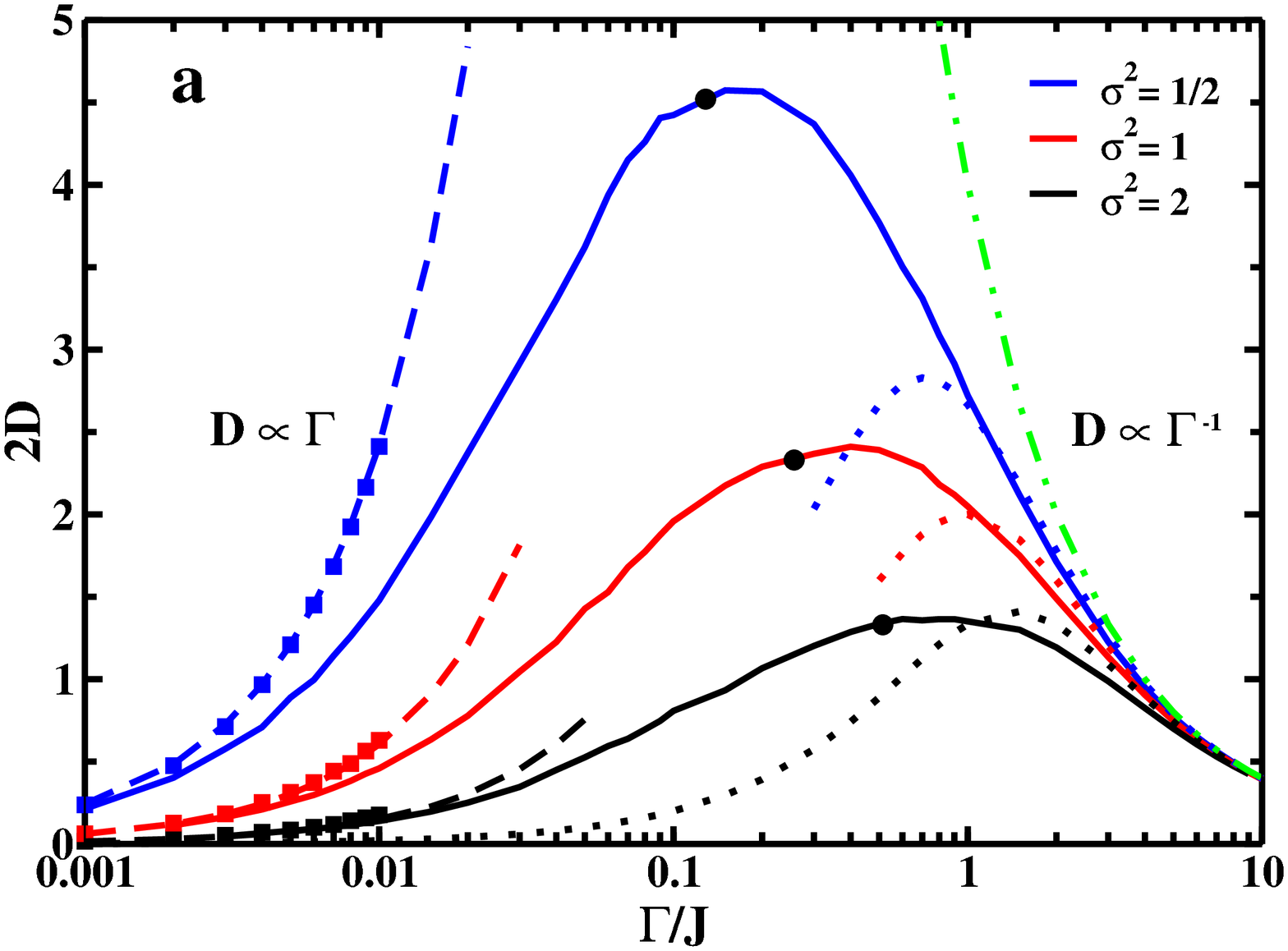}
       \label{fig:diffusion}     
    }
   \end{subfigure}
   \begin{subfigure}{
       \includegraphics*[width=0.5\textwidth]{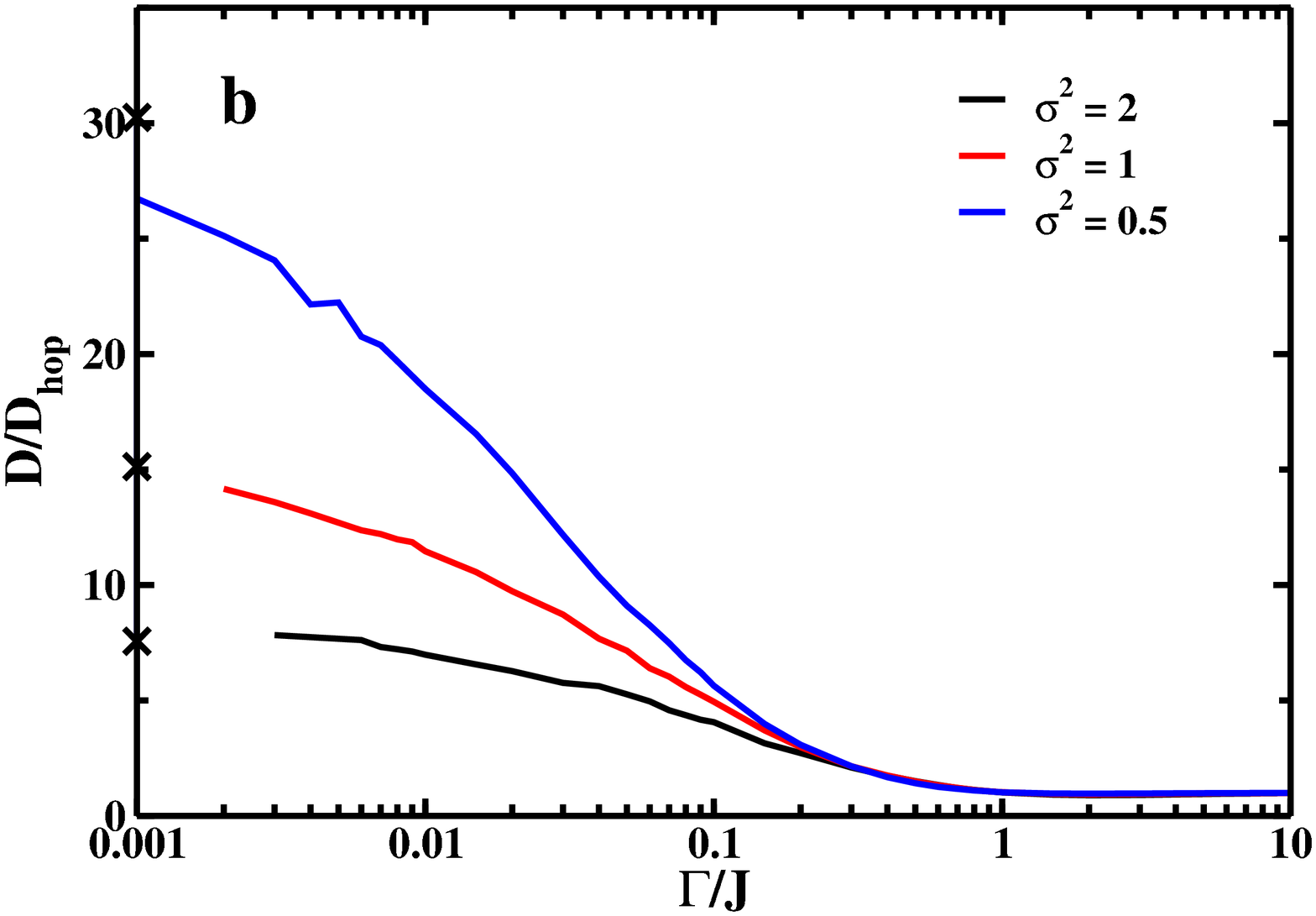}
       \label{fig:speedup}
    }
   \end{subfigure}
\caption{
   (a)
   Diffusion constants calculated as a function of the dephasing rate.
   The solid blue, red, and black lines are the exact numerical simulations 
   of~\eref{eq:HakenStrobl} 
   for $\sigma^2/J^2 = 1/2$, $1$, and $2$, respectively.
   The filled squares at small dephasing display the results of
   the secular Redfield equation and the dashed lines provide the 
   corresponding analytic results of~\eref{eq:redfield_scaling}.
   At large dephasing, the dotted lines depict the 
   classical hopping estimates of~\eref{eq:hopping} and the dot-dashed
   line is the exact result of~\eref{eq:Reineker} for $\sigma^2=0$.
   The solid black dots indicate the estimate for location of the 
   optimal dephasing rate provided in~\eref{eq:D_max}.
   (b) The ratio of the numerically exact diffusion constant in (a) 
   to the classical hopping result of~\eref{eq:hopping}.
   The cross-symbols on the ordinate indicate the limiting values
   given by $D_{\rm coh}/D_{\rm hop}$ in~\eref{eq:speedup}.
}
   \end{center}
\end{figure}

The diffusion coefficients calculated from the above approaches are 
shown in figure~\ref{fig:diffusion} as a function of the dephasing rate.
In agreement with the discussions of the previous section, 
When $\Gamma/J \gg 1$ the exact numerical results agree with 
the prediction of the classical hopping model in~\eref{eq:hopping}
(dotted lines). 
At even larger values of the dephasing, the diffusion constant becomes
independent of the disorder and the results coincide with estimate 
of~\eref{eq:Reineker} for the homogeneous chain (dot-dashed line).
However, one notices that as the disorder decreases and the dynamics
become more coherent the classical hopping approximation quickly deteriorates
in predicting both the magnitude of the diffusion constant as well as the
location of the maximum.
For sufficiently weak dephasing, $\Gamma/J \ll 1$, 
the exact numerical results are in agreement with the numerical calculations 
of the Redfield equation (square symbols in figure~\ref{fig:diffusion}).
The latter are seen to be in excellent agreement with the scaling relation 
of~\eref{eq:redfield_scaling} (dashed lines),
with the localization length, $\xi$, determined from 
the IPR calculations in figure~\ref{fig:ipr}.

In order to further explore the role of coherence in the dynamics, 
figure~\ref{fig:speedup} displays the ratio of the exact numerical diffusion
coefficients shown in figure~\ref{fig:diffusion} to the classical hopping
prediction, $D/D_{\rm hop}$.
For $\Gamma/J>1$, the exact and classical hopping results are 
approximately equivalent since an coherences generated during the time
evolution of the populations are quickly destroyed by the strong dephasing.
However, as the dephasing decreases the dynamics become more coherent
leading to a significant increase in the ratio of $D/D_{\rm hop}$.
In the asymptotic regime where $\Gamma/J\ll 1$, the exact numerical results
approach the limiting values given in~\eref{eq:speedup} (cross symbols) 
which are determined only by the Anderson localization length of the 
static system Hamiltonian. 
As discussed above, the transition from coherent dynamics in 
the weak dephasing regime to the classical hopping dynamics
is due to the fact that in addition to the static disorder, 
the thermal environment can also serve to localize the system.
That is, the true localization length is a function of both
$\sigma$ and $\Gamma$.
In the limit $\Gamma=0$, the localization is entirely due to Anderson
localization whereas for $\Gamma/J\gg 1$ the localization is largely
due to the dephasing.
Indeed, the results in figure~\ref{fig:speedup} may be interpreted as a measure
of the true localization length in the system.

\begin{figure}
   \begin{subfigure}{
         \includegraphics*[width=0.5\textwidth]{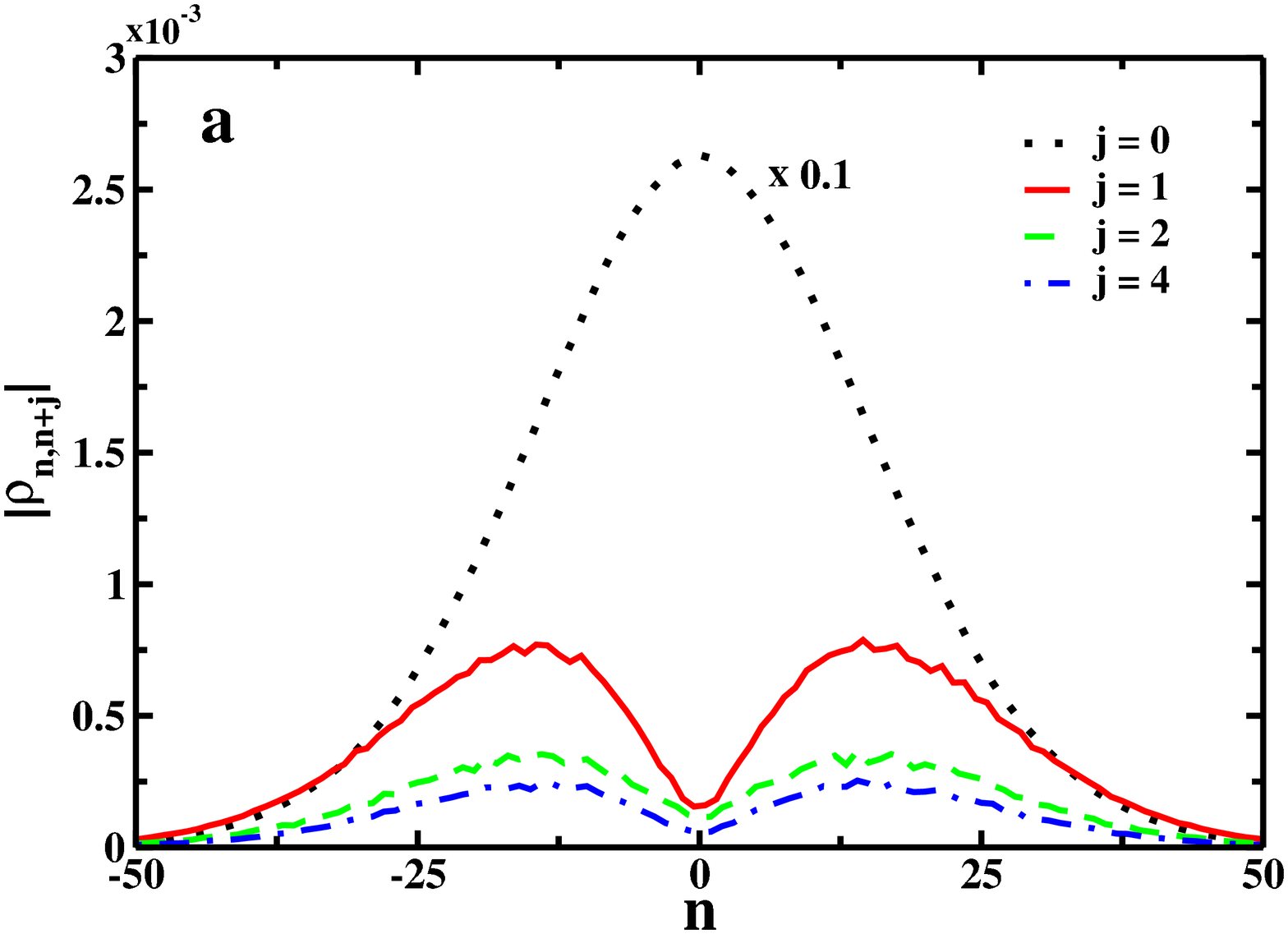}
       \label{fig:rho}     
   }
   \end{subfigure}
   \begin{subfigure}{
         \includegraphics*[width=0.5\textwidth]{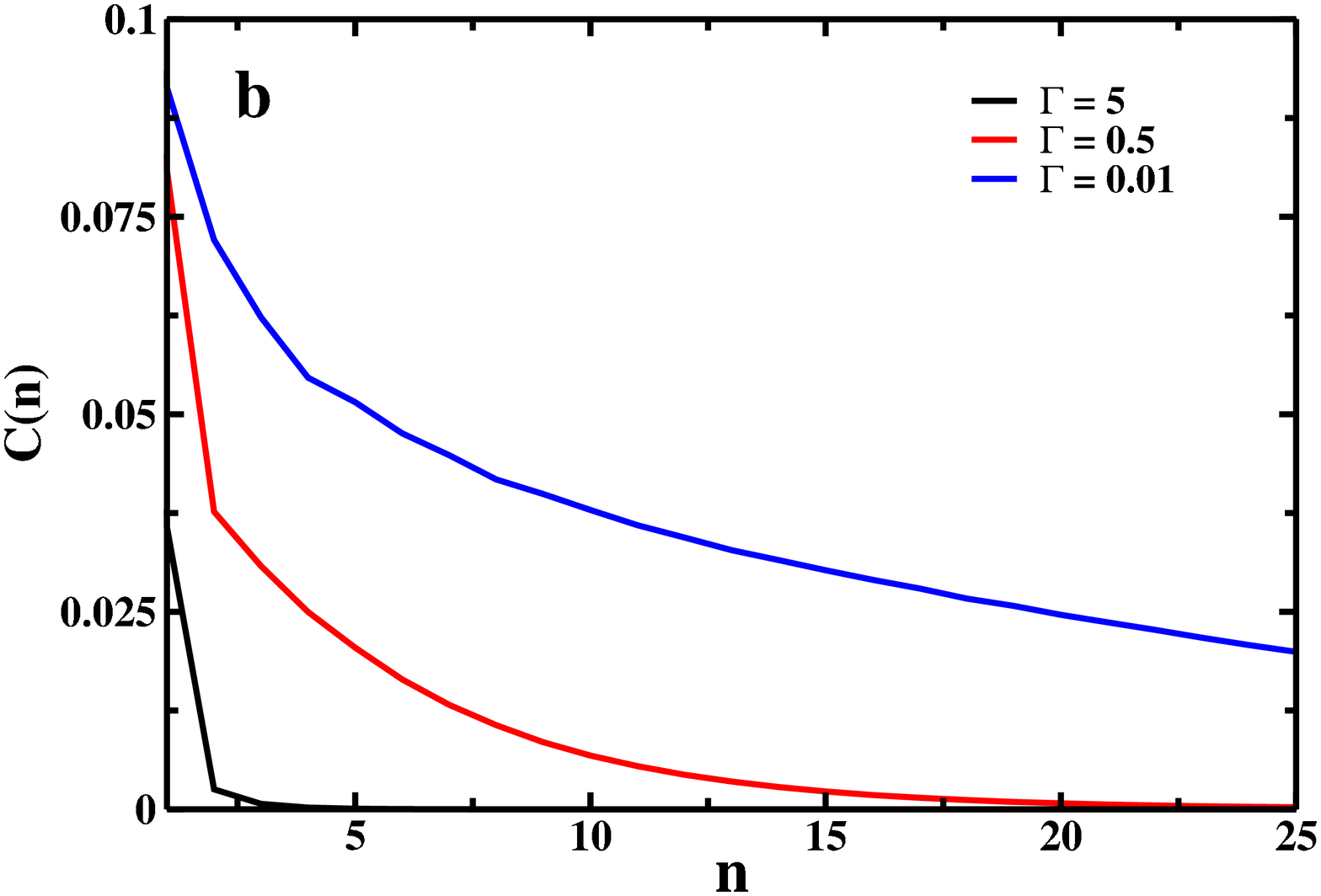}
       \label{fig:coherences}     
   }
   \end{subfigure}
\caption{
   (a) The absolute value of the density matrix elements computed in the 
   steady state for the parameters $\Gamma/J = 0.5$ and $\sigma/J=1$. 
   The various values of $j$ represent the $j$-th superdiagonal of the density
   matrix.
   The populations $(j=0)$ have been scaled by a factor of $0.1$
   in order to reside on the same scale as the coherences ($j=1,2$, and $4$). 
   (b)
   The mean coherences present in the steady state density matrix 
   as determined from~\eref{eq:coherence} with $\sigma/J=1$.
}
\end{figure}

As mentioned, the dephasing can play an equally important role to the static
disorder in localizing the system. 
It would be highly desirable to be able to quantify the localization that
occurs due to the dephasing. 
This is generally affected through computations of the equilibrium reduced 
density matrix~\cite{moix12}.
However, the equilibrium state of systems described by the HSR model is
somewhat trivial due to the infinite temperature bath. 
Each site is equally populated and the system is completely localized
in the site basis.
Based upon this observation one would expect the classical hopping description
to always be applicable, which is clearly not the case as shown above.
The failure of this approach lies in the realization that there is an inherent 
difference between the equilibrium state in a large, but finite chain and 
the nonequilibrium steady-state of an infinite chain where the populations 
are continuously evolving.
If one considers the distribution of populations, 
then tails of the distribution represent the sites 
where the population is spreading to new sites in the chain. 
In this region coherences are constantly being created and destroyed.
The spatial extent of these coherences also reaches a steady state,
and it is this coherence length that is responsible for the deviations from 
the classical hopping result. 
The interior of the probability distribution, however, equilibrates.
In order to demonstrate this point more clearly, figure~\ref{fig:rho} displays 
slices through a snapshot of the steady-state density matrix
taken in the regime where an accurate estimate of the diffusion coefficient 
is possible.
One sees that the populations (dotted line) display the Gaussian 
profile expected of a diffusion process.
The coherences are substantial only in the tails of the population
distribution, while the interior thermalizes and becomes diagonal in the 
local basis.
In an infinite chain, this process continues indefinitely. 
At longer times, the coherences in the center of the chain will be further 
depleted while new coherences will be created in the extremities 
as the populations evolve.

In order to quantify the role of dephasing in localizing the system, 
it is necessary to characterize the amount of coherence present
in the density matrix in the diffusive, steady-state regime.
Unfortunately, as seen above there is no way to unambiguously determine 
this coherence length.
As a simple qualitative measure we will use the metric,
\begin{equation}
   C(n) = \sum_{i}\left|\overline{\rho_{i,i+n}} \right|
   \;, \label{eq:coherence}
\end{equation}
computed from the steady-state density matrix in the long-time limit where a 
reliable estimate of the diffusion constant is possible.
The correlation function $C(n)$ is simply the sum of the individual
superdiagonals of the density matrix displayed in figure~\ref{fig:rho}.
As can be seen in figure~\ref{fig:coherences}, both the magnitude
and spatial extent of $C(n)$ clearly depend on the dephasing.
For $\Gamma/J\gg1$ the dephasing may completely localize the system
in the site basis regardless of the disorder.
Unsurprisingly, in this regime the classical hopping approximation to the 
transport is highly accurate as seen above in figure~\ref{fig:diffusion}.
In the opposite regime, the dephasing has less influence on localizing 
the system, and the extent of the wavefunctions is predominantly determined 
by the Anderson localization due to static disorder.
Optimal diffusion occurs in the intermediate regime, where both the noise 
and disorder play a role in the localization of the wavefunctions.

\subsubsection{Improved diffusion estimates}

\begin{figure}
   \begin{center}
   \begin{subfigure}{
         \includegraphics*[width=0.5\textwidth]{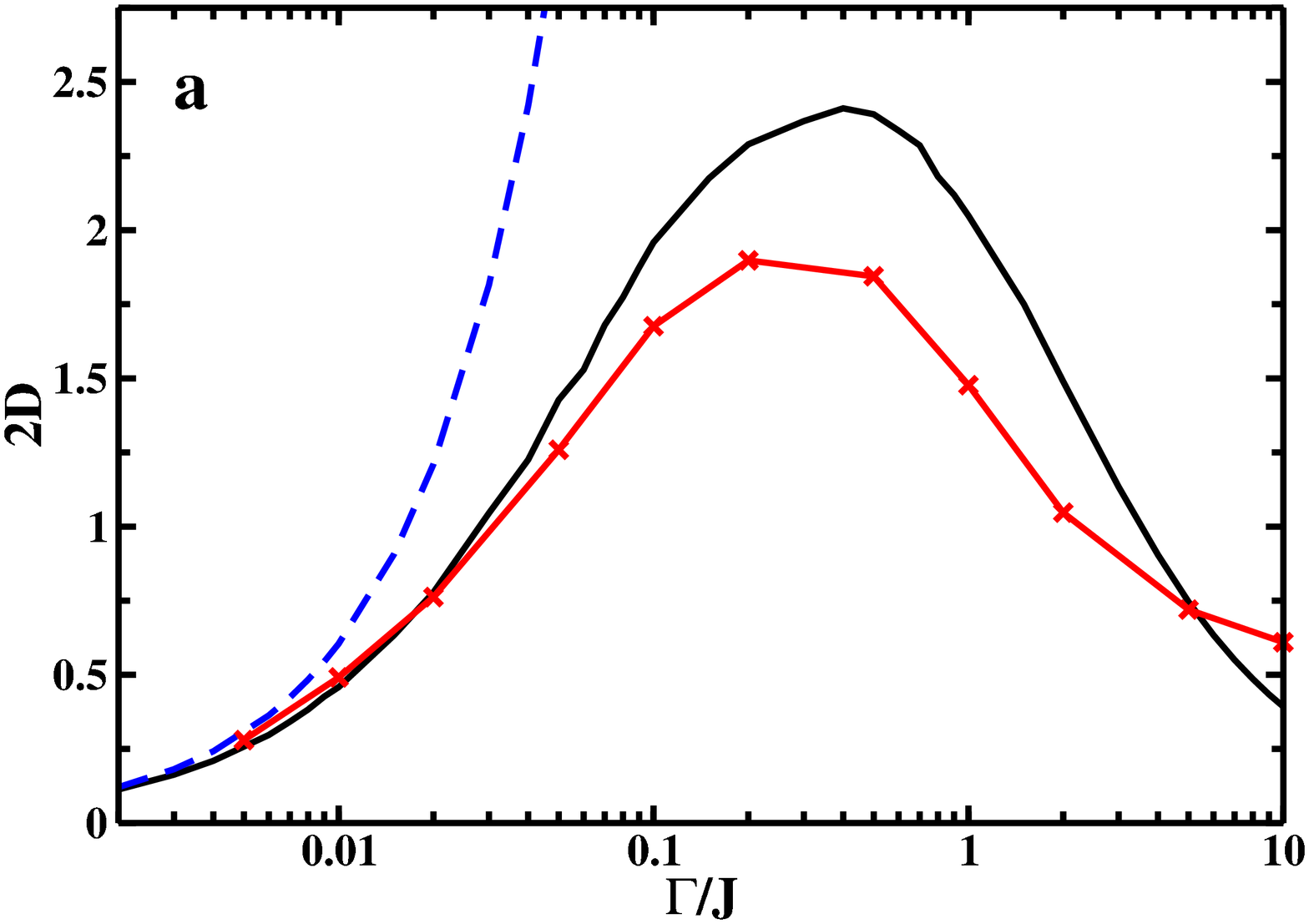}
       \label{fig:gamma_loc} 
   }
   \end{subfigure}
   \begin{subfigure}{
         \includegraphics*[width=0.5\textwidth]{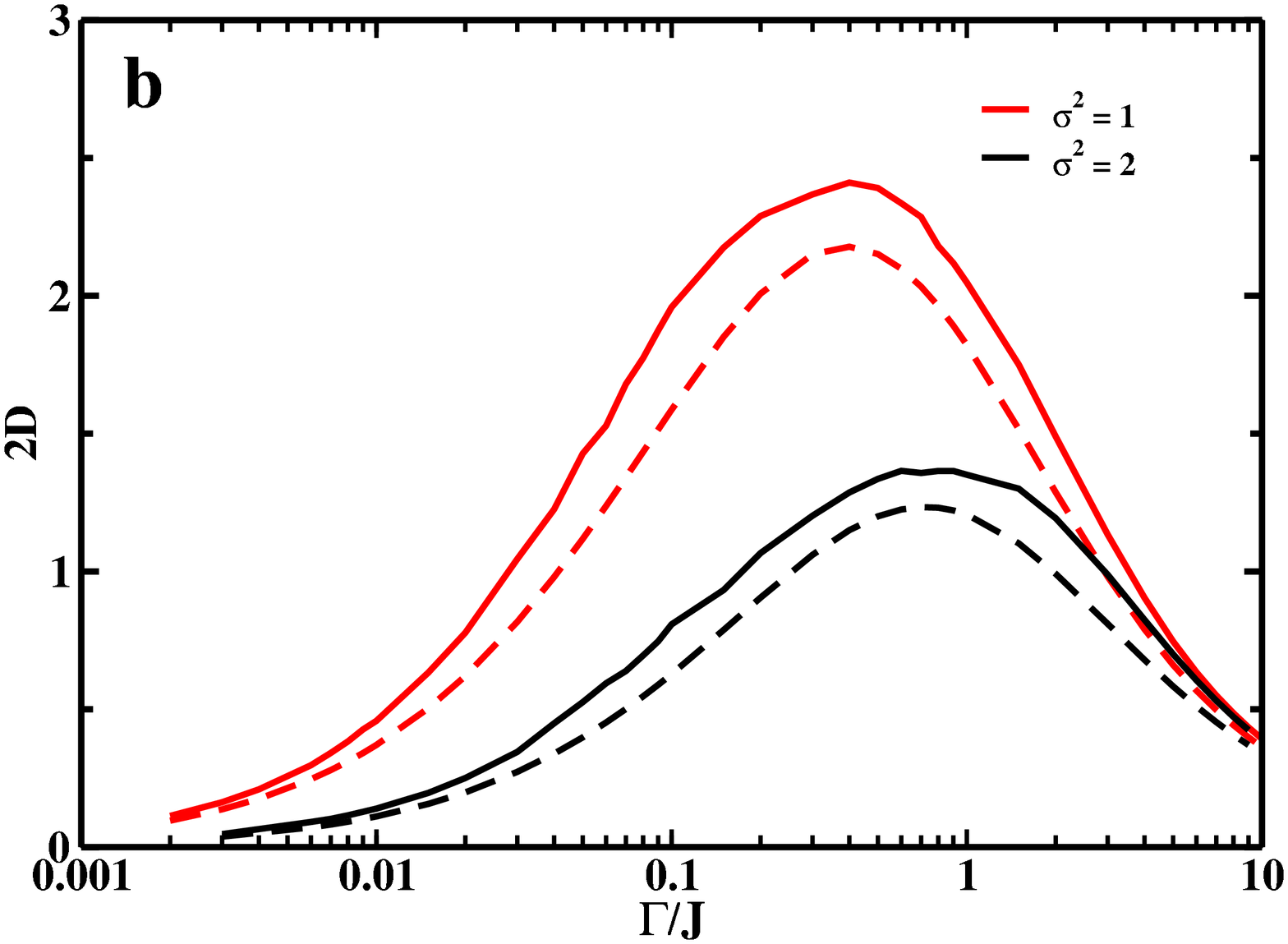}
       \label{fig:2d_fit}
    }
   \end{subfigure}
   \caption{
   (b) The exact numerical results for the diffusion constant
   for $\sigma/J=1$ reproduced from figure~\ref{fig:diffusion}
   along with the weak dephasing result from~\eref{eq:redfield_scaling}
   (blue dashed line).
   The solid (red) line with cross symbols denote the improved
   scaling relation of~\eref{eq:redfield_thermal}.
      (b) 
      The exact numerical results for the diffusions coefficients 
      from figure~\ref{fig:diffusion} (solid lines) 
      computed at $\sigma^2/J^2 = 1$ (red) and $2$ (black).
      The corresponding approximate results of~\eref{eq:thouless} are
      displayed as the respective dashed lines.
   }\label{fig:2D}
   \end{center}
\end{figure}

It is interesting to explore the question if the thermal
localization can be used to extend the regime of validity of the scaling 
behavior obtained from the Redfield equation in~\eref{eq:redfield_scaling}.
To this end, we propose a simple ansatz for a dephasing-dependent correction 
to localization length, $\tilde \xi(\Gamma) = f(\Gamma)\xi$, 
with $0 \le f(\Gamma) \le 1$,
which leads to the corresponding diffusion constant,
\begin{equation}
   D_{\rm coh,\Gamma} = \Gamma \tilde\xi(\Gamma)^2
   \;.
   \label{eq:redfield_thermal}
\end{equation}
The behavior of $f(\Gamma)$ can be determined by noting that 
for disordered systems, $C(n)$ falls off exponentially for large $n$ 
regardless of the dephasing.
That is, $C(n)\sim e^{-n/L_{\Gamma}}$ with the 
characteristic length scale denoted by $L_\Gamma$.
Therefore, it is possible to define the correction factor 
$f(\Gamma) = L_{\Gamma=0}/L_\Gamma$, where $L_{\Gamma=0}$ is obtained
in the dephasing-free, Anderson localization limit.
From the steady-state density matrix, we compute $C(n)$ and 
then fit the exponential decay at large $n$ to obtain $f(\Gamma)$.
The results of this procedure for computing the dephasing-corrected
diffusion coefficient are displayed in figure~\ref{fig:gamma_loc}.
For small dephasing rates, the modified diffusion coefficient 
corrects the overestimation of~\eref{eq:redfield_scaling} and 
even predicts the turnover seen in the exact results.
The agreement with the exact numerical results is of course only qualitative, 
but one could easily devise a more sophisticated approach by
demanding, for example, that~\eref{eq:redfield_thermal}
also reproduces the correct large dephasing limit.
However the main point of this argument is to provide and 
intuitive illustration of the influence of dephasing and disorder 
on the transport dynamics discussed above.

While perhaps useful, 
it should be noted that this procedure for correcting the weak dephasing 
scaling limit is not a constructive approach.
That is, computing the dephasing-induced localization length
requires the steady-state density matrix,
which alone is sufficient to determine the exact diffusion constant.
However, a more direct and accurate approach to characterize the thermal 
localization length could prove a useful route to a uniform approximation of
the  diffusion constant.

Although the above discussion provides a physically motivated uniform 
approximating to the diffusion coefficient,  
a simple alternative has been proposed by Thouless and 
Kirkpatrick~\cite{thouless81}.
They cleverly suggest to use an interpolating formula,
\begin{equation}
   D_{\rm interp} \approx 
                     \left[ \left(\frac{2J^2}{\Gamma + \sigma/2}\right)^{-1/2} 
                     + \left(\Gamma \xi^2 \right)^{-1/2}\right]^{-2} 
                     \;, \label{eq:thouless}
\end{equation}
which reduces to the correct limits in both the strong and weak dephasing 
regimes, and reasonably bridges the intermediate region.
Figure~\ref{fig:2d_fit} displays the exact numerical results for the 
diffusion constant computed previously in figure~\ref{fig:diffusion} with
$\sigma^2/J^2=1$ and $2$ (solid lines),
along with the corresponding approximation from~\eref{eq:thouless} 
(dashed lines).
This approach performs well in one-dimensional systems discussed
here, although in our preliminary studies of two-dimensional systems
this simple interpolation appears to be less satisfactory.

\subsection{Non-diffusive dynamics}
\label{sec:msd}


\begin{figure}
\begin{center}
   \includegraphics*[width=0.5\textwidth]{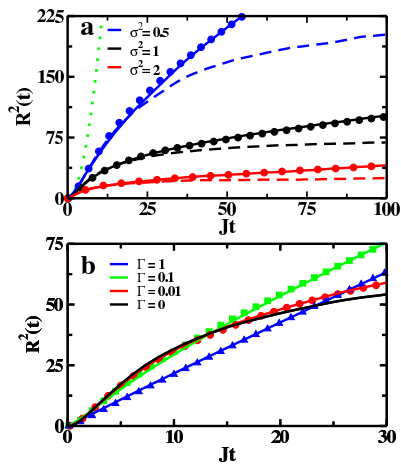}
\caption{
   In (a), the dashed blue, red, and black lines depict the MSD for 
   $\Gamma=0$, and  $\sigma^2 = 1/2$, $1$, and $2$.
   This case is exactly the bounded dynamics due to Anderson localization
   and provides a lower bound to the transport.
   The solid lines depict the corresponding results with $\Gamma=0.01$.
   The green dotted line is the exact result of~\eref{eq:msd} 
   for $\sigma=0$. 
   This case represents free diffusion in the homogeneous chain and
   defines the upper bound to the transport rate.
   Figure~(b) displays the MSD at very short time for $\sigma^2=1$.
   The black, red, green, and blue lines correspond to dephasing rates of 
   $\Gamma=0$, $0.01$, $0.1$, and $1$.
   The symbols are exact numerical results and
   the solid lines are the fits to~\eref{eq:msd_fit} 
   with $\eta=0.84$, $0.2$ and, $0.04$.
}\label{fig:msd}
\end{center}
\end{figure}

While the behavior of the diffusion coefficient is interesting,
it is also useful to understand the dynamical properties of 
noisy, disordered systems.
In figure~\ref{fig:msd}(a) the initial MSD
is shown for short times ($\Gamma t=1$). 
For $\sigma=0$ but finite dephasing, the exact MSD 
is given by~\eref{eq:msd}.
Over the time scale shown, this result (dotted line) leads to 
purely ballistic, free-particle motion,
and substantially overestimates the transport.
For $\Gamma=0$, but finite disorder (dashed lines), 
the motion is again described by that of a free particle, 
although now in a disordered environment.
In this case, the dynamics are also ballistic, but only 
until the localization length is reached at which point the transport stops. 
This behavior is similar to that of a particle in a box where the 
size of the box is determined by the Anderson localization length.
The limit of $\Gamma=0$ provides a lower bounder to the transport
whereas the homogeneous chain limit of $\sigma=0$ provides an upper bound.

The solid lines in figure~\ref{fig:msd}(a) display the 
results for the corresponding disordered system with the addition
of weak dephasing ($\Gamma/J=0.01$).
As can be seen, the coupling to the high-temperature environment 
allows the wavepacket to overcome the localization barriers, 
eventually leading to diffusion.
Additionally, for short times the MSD 
essentially mirrors that of the corresponding $\Gamma=0$ results.
That is, for weak dephasing, the initial dynamics can be qualitatively
described by those of a free particle in a disordered chain.
Consistent with the arguments that led to the estimate for the
diffusion constant in~\eref{eq:redfield_scaling}, there are two time 
scales present in the weak dephasing limit. 
The fast time scale is described by the (ballistic)
free particle expansion in the Anderson localized chain.
Over a longer time scale, of order $\Gamma^{-1}$, 
the diffusive transport becomes dominant due to role of the dephasing.
It is interesting to note the possibility that for sufficiently small
dephasing the non-diffusive motion could persist over a
timescale such that it is comparable with exciton dissociation 
or recombination.

In figure~\ref{fig:msd}(b), the role of the dephasing rate 
on the short-time, free-particle behavior is shown.
For smaller dephasing rates, the dynamics are essentially identical to
the $\Gamma=0$ result over the time scale shown.
At very short-times, the free particle motion ($\Gamma=0$) 
provides an upper bound to the rate of spreading.
As the dephasing rate increases, the departure of the respective 
MSD from the $\Gamma=0$ result occurs sooner.
Similar transient localization effects have been observed in
semiclassical simulations of the transport of organic 
semiconductors and quasicrystals~\cite{ciuchi11,trambly_de_laissardiere06}.
Based on these observations, one can approximate the MSD 
at any dephasing rate as a linear combination of the $\Gamma=0$, free-particle 
result and the long-time linear diffusion, 
\begin{equation}
\overline{\left \langle R(t)^2 \right \rangle} \approx 
2 dD t + \eta(\sigma,\Gamma)\overline{\left \langle R(t)^2 \right \rangle}_{\Gamma=0},
   \label{eq:msd_fit}
\end{equation}
where the parameter, $0 \le \eta\le 1$. 
As the dephasing rate increases, the prefactor, $\eta$, decreases.
The symbols in figure~\ref{fig:msd}(b) display the exact numerical 
results and the solid lines are the corresponding fits to~\eref{eq:msd_fit}.
This fit becomes quantitative in the small dephasing regime 
where $\eta\sim 1$.
In the general case,~\eref{eq:msd_fit} accurately 
describes the dynamics at both short and long times, 
with a small discrepancy in the intermediate regime.

\begin{figure}
\begin{center}
   \includegraphics*[width=0.5\textwidth]{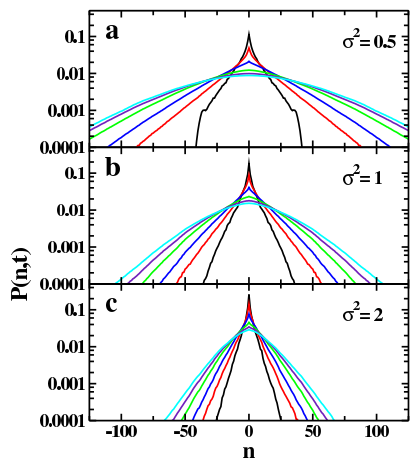}
\caption{
   The probability distributions at times 
   $\Gamma t = 0.2$ (black), $2$ (red), $5$ (blue), $10$ (green), 
   $15$ (purple) and $20$ (cyan) calculated with $\Gamma=0.01$.
   Panels (a), (b), and (c) correspond to $\sigma^2=0.5$, $1$, and $2$, 
   respectively.
}\label{fig:dists}
\end{center}
\end{figure}

Figure~\ref{fig:dists} displays the average probability distributions
of the site populations at
$\Gamma/J=0.01$ and $\sigma^2/J^2=0.5$ (a), $1$ (b), and $2$ (c) 
for an initial excitation located at a single site in the center 
of the disordered chain.
In all three cases the tails of the distribution functions appear 
exponential at short times and transition to Gaussian at long times.
In driven disordered systems~\cite{yamada99}, it was proposed that the 
probability distribution at any fixed time can be fit as
$
   P(n) \propto e^{-\phi\left|n\right|^\gamma} 
   $,
where the scaling exponent, $1 \le \gamma  \le 2$ and $\phi \ge 0$.
Physically, this expression arises from the observation that 
at short times, all of the eigenstates are exponentially localized,
which leads to a corresponding exponential probability 
distribution for the populations ($\gamma=1$).
At long times, the motion is diffusive implying a 
Gaussian probability distribution with $\gamma=2$.
For the HSR model, this picture is only qualitatively correct.
It clearly can not capture the persistent sharp peak at the origin that arises 
from the initial excitation or the wavefront of the initial 
ballistic spreading seen at $\Gamma t = 0.2$ for $\sigma^2=0.5$.

\subsection{Absorption spectra}

\begin{figure}
   \begin{center}
   \begin{subfigure}{
         \includegraphics*[width=0.5\textwidth]{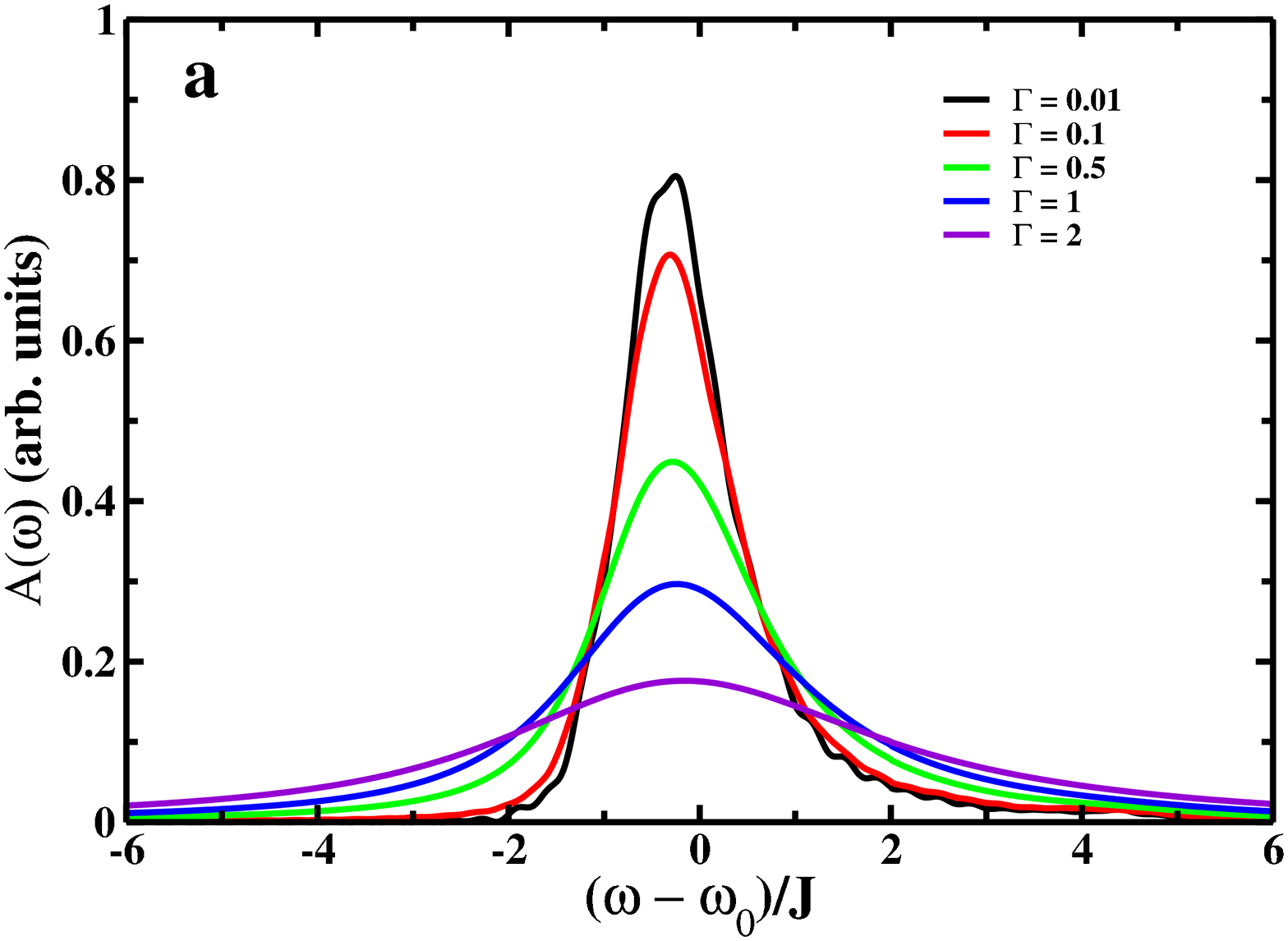}
      \label{fig:spectrum}
   } 
   \end{subfigure}
   \begin{subfigure}{
         \includegraphics*[width=0.5\textwidth]{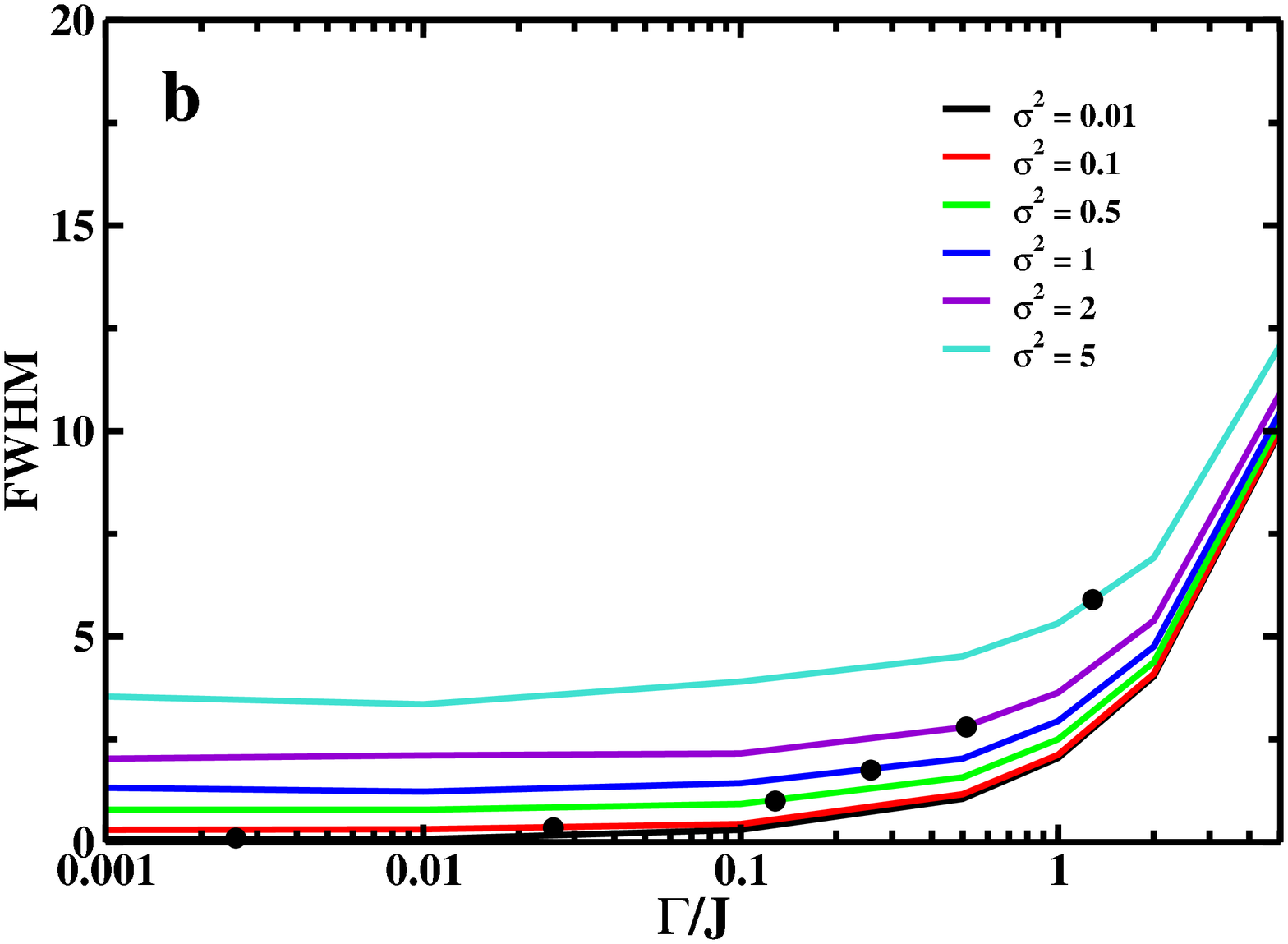}
      \label{fig:fwhm}
   }
   \end{subfigure}
   \caption{
      (a) Absorption spectra of a disordered linear chain calculated 
      at $\sigma/J=1$.
      (b) Full width at half maximum as a function of the 
      dephasing rate for various values of the disorder.
      The solid dots display the value of the dephasing
      for which the transport is maximized from~\eref{eq:D_max}.
   }
   \end{center}
\end{figure}

From the above results, it is clear that the diffusion coefficient 
is strongly influenced by the amount of disorder present in the system.
While reducing the amount of disorder will improve the transport
properties, this is often not the only material characteristic that
is important for the overall device performance.
For instance, in the case of organic photovoltaics the organic semiconductor
should ideally possess not only a high mobility but also a broad absorption 
spectrum in order to capture the largest portion of the solar spectrum. 
Therefore in order to understand the interplay of the transport properties
and absorption characteristics, 
figure~\ref{fig:spectrum} displays the effect of the
dephasing on the absorption spectrum computed 
from~\eref{eq:absorption} with $\sigma^2/J^2=1$.
In contrast to the diffusion coefficient, the dipole-dipole correlation 
function decays rapidly to zero on a timescale on the order of $\Gamma^{-1}$.
That is, the absorption spectrum is determined by the short-time,
non-diffusive dynamics of the wavepacket (cf.~figure~\ref{fig:msd})
rather than the long-time linear behavior of the MSD.
In figure~\ref{fig:spectrum}, the spectra computed at $\Gamma/J=0.01$ 
is indistinguishable from the spectrum 
calculated with disorder only (see~\eref{eq:disorder_spectra})
and is governed only by inhomogeneous broadening mechanisms.
In this case, the lineshape displays the characteristic asymmetric profile 
commonly observed in the low-temperature spectra of excitonic 
systems --a Gaussian profile for $\omega < \omega_0$ 
and Lorentzian lineshape for $\omega > \omega_0$~\cite{klafter78, fidder91a}. 
Only at very large disorder, $\sigma/J\gg 1$, does the inhomogeneously
broadened spectrum become purely Gaussian.
As the dephasing is increased, the spectra are broadened into a Lorentzian 
lineshape indicative of systems dominated by homogeneous broadening.
A detailed discussion of the absorption spectra and the associated
lineshape is provided in~\ref{appendix:absorption}.

In order to investigate the relationship between spectral properties and
transport properties in more detail, it is useful to 
analyze the full width at half maximum (FWHM) of the absorption spectrum 
as a function of the dephasing rate and disorder as shown 
in figure~\ref{fig:fwhm}.
As discussed previously at small dephasing, the linewidth is determined solely
by inhomogeneous broadening mechanisms which leads to the plateau in
the FWHM for $\Gamma/J \ll 1$.
In the opposite regime of large dephasing, the spectra are purely Lorentzian
with a linewidth that scales linearly with the dephasing rate
as discussed in~\ref{appendix:absorption}.
The solid dots in figure~\ref{fig:fwhm} denote the value of the dephasing 
for which the transport is maximal as predicted from the scaling relation
of~\eref{eq:D_max}.
At small disorder
optimal transport occurs only at 
vanishing dephasing rates where the FWHM is almost zero indicating an
unfavorable, narrow absorption lineshape.
In the opposite limit of very large dephasing the spectrum is quite broad, 
but the diffusion rate is also negligible (see~\eref{eq:Reineker}),
which is also not optimal.
As a result, engineering a material with both favorable transport properties 
and a absorption profile requires a delicate balance of 
the disorder and dephasing.

\section{Applications}\label{sec:applications}
The dephasing rate in the HSR model may be obtained from a 
microscopic description of the environment 
that is characterized by a thermal bath of harmonic oscillators.
By taking the high temperature limit of the bath correlation function
one obtains the simple relationship,
$\Gamma = 2 k_{\rm b} T \lambda/\omega_d$, 
where $\lambda$ is the reorganization energy, $\omega_d$ is the associated
Debye frequency of the material,
and $T$ is the temperature~\cite{jayannavar88,wu10,valleau12,lloyd11}.
Substituting this expression into~\eref{eq:D_max} leads to
a relationship for the temperature that maximizes
the diffusion coefficient,
\begin{equation}
   k_{\rm b} T_{\rm max} \propto \frac{J \omega_d}{2 \lambda \xi}
   \;.
   \label{eq:T_max}
\end{equation}
While a non-monotonic temperature dependence of the transport properties
has indeed been observed experimentally in several molecular crystal thin 
films such as  rubrene and perylene~\cite{karl99, podzorov04},
a more accurate treatment of the thermal environment is required in 
order to conclusively confirm this interesting observation.
Preliminary numerical calculations for a disordered system
weakly coupled to a quantum harmonic bath have demonstrated that 
that the diffusion constant does indeed exhibit a maximum as a function
of temperature as predicted here~\cite{zhong13b}.
However, the prediction of~\eref{eq:T_max} has been shown to
be only qualitatively correct. 

In addition to the non-monotonic temperature dependence of the transport
properties, recent experiments on the effect of disorder 
on the conductivity of the conducting polymer, polyaniline, also clearly 
illustrate the principles outlined in this work~\cite{lee06}.
With conventional processing techniques, samples of polyaniline
are highly disordered.
This is manifested in experimental measurements 
by the appearance of a minimum in the resistivity as a function of temperature,
and at temperatures below this point the resistivity increases rapidly 
due to disorder-induced localization.
Through improved sample processing methods however, 
Lee et al.~were able to reduce the amount of disorder present 
in polyaniline~\cite{lee06}.
They demonstrated that the resistivity decreases uniformly as
the sample quality improves, and additionally, 
the location of the minimum shifts to lower temperatures.
This behavior is consistent with that seen in figure~\ref{fig:diffusion}.
By reducing the amount of disorder even further, the resistivity minimum 
disappears completely which has been interpreted as the 
the onset of metallic behavior in polyaniline~\cite{lee06}.

Finally as a simple predictive example,
one may estimate the temperature at which the transport
is maximized in arrays of the natural light harvesting system, 
LH2~\cite{escalante10}.
A one-dimensional array of these photosynthetic complexes is relatively well 
characterized by the parameters,
$\lambda=\sigma=200$ cm$^{-1}$, $\omega_d=50$ cm$^{-1}$, and an inter-ring 
coupling of $J=50$ cm$^{-1}$.
With these values,~\eref{eq:T_max} predicts a peak in the transport at 
$k_b T \approx 25$ cm$^{-1}$.
However, as mentioned above, in order to provide a definitive estimate, it will
be essential to develop a more accurate description of the environment
beyond the HSR model.

\section{Conclusions}

In conclusion,
we have provided exact numerical results for the transport properties
of infinite one-dimensional disordered systems coupled to a classical
thermal environment.
The limiting behavior at weak and strong dephasing, elucidated
through the development of master equations in the respective limits, 
imply the existence of a maximal diffusion rate at intermediate dephasing.
This prediction and the respective scaling relations have been confirmed
numerically.
While the results have been obtained within the HSR model, the scaling
relations are generic and their generalization to systems described by a more
realistic environment should be possible.
Additionally we have shown that in the weak dephasing limit, 
the coherent quantum transport leads to 
diffusion rates that are much larger than would be predicted classically
with an enhancement factor that is proportional to the localization length.
The numerical results presented in section~\ref{sec:msd} on the 
dynamical properties of the transport demonstrate that 
the MSD in the weak-dephasing regime may be described 
as a sum of the short-time, free particle motion in the Anderson localized
chain and the long-time linear diffusion. 
The former behavior appears as a significant non-Gaussian contribution
during the evolution of the probability distributions of the site populations.
Finally the influence of the noise and disorder on the spectral 
lineshapes and the relation between the absorption spectrum and 
the optimal transport regime 
have been presented in the context of organic photovoltaic
materials.
While decreasing the amount of disorder 
improves the transport properties, it also leads to an 
undesirable narrowing of the absorption spectra.
The need for both a broad spectrum and favorable
transport characteristics requires a delicate balance of the
disorder and dephasing.

The work presented here lays the foundation for a series of forthcoming 
results.
As mentioned briefly in section~\ref{sec:applications}, the extension
of the HSR environment to a quantum environment is currently
being explored through the use of recently developed 
non-Markovian stochastic Schr{\"o}dinger equations~\cite{zhong13}.
Preliminary results display finite zero-temperature transport
as well as an optimal diffusion rate as a function of 
temperature~\cite{zhong13b}.
An additional topic of current focus is extending the model
to higher dimensions and using more realistic non-nearest neighbor 
interactions to describe the electronic couplings.
As has been recently reported, disordered systems with long-range
interactions can develop a mobility edge and undergo an Anderson 
transition~\cite{rodriguez03}.
This has a profound impact on the transport properties and absorption
characteristics.
Finally, we are also exploring an extension of the HSR model 
in which the classical environment is modified to include non-Markovian 
effects describing the coupling of the system to under-damped vibrations.
Both the relaxation time of the bath as well as the frequency of the
vibrational modes have been seen to have a important influence on
the transport properties.
These interesting preliminary results are topics of ongoing work.

\section{Acknowledgments}

This work was supported by 
the NSF (Grant No. CHE-1112825) and 
DARPA (Grant No.~N99001-10-1-4063).
J.~Moix and M.~Khasin are supported by the Center for Excitonics,
an Energy Frontier Research Center funded by the US Department of Energy,
Office of Science, Office of Basic Energy Sciences under Award 
No.~DE-SC0001088.
J.~Moix acknowledges a generous allocation of computing 
time from the Extreme Science and Engineering Discovery Environment
(XSEDE), which is supported by the NSF (Grant No.~OCI-1053575).

\appendix


\section{Absorption Spectrum}\label{appendix:absorption}

Herein exact expressions for the absorption spectrum in the presence of
dephasing and disorder are presented~\cite{mukamel85}.
In the absence of the bath, the absorption spectra is given 
by Fermi's Golden Rule~\cite{fidder91a},
\begin{equation}
   A(\omega) = \frac{1}{N}\sum_{\kappa} \left|\mu_\kappa \right|^2 
               \delta(\omega-\omega_{\kappa}) 
   \;, \label{eq:disorder_spectra}
\end{equation}
where the dipole moments are given by
\begin{equation}
   \mu_\kappa = \mu_0 \sum_n \left\langle \kappa | n \right \rangle
   \;.  \label{eq:dipole}
\end{equation}
As before, Greek indices denote the basis of eigenstates of $H_s$ 
with the corresponding eigenvalues, $\omega_\kappa$.
Henceforth, we assume that all of the molecules in the system
are identical and possess the same dipole moment, $\mu_0 = 1$. 

In the presence of the environment the absorption
spectra is more conveniently obtained in the time domain 
from the dipole autocorrelation function (cf.~\eref{eq:absorption}),
\begin{equation}
   \left \langle \mu(t) \mu(0) \right \rangle
   = {\rm Tr}\left[\mu e^{-i L t} \mu  \rho_g  \right]
   \;,
\end{equation}
where $\rho_g=|0\rangle\langle 0|$ is the ground state density matrix,
and $L$ is the Liouville operator for the total system and bath.  
Defining the initial state $\tilde \rho = \mu \rho_g$, 
its corresponding time evolution is given by the Liouville equation,
\begin{equation}
   \dot {\tilde \rho} = -i \left[ H_s, \tilde \rho\right] 
   - \frac{\Gamma}{2} \sum_{n=1} \left[V_n,\left[V_n,\tilde \rho\right]\right] 
   \;.
\end{equation}

Since the ground state is not coupled
to the single excitation manifold or the bath, 
the equation of motion for the density matrix reduces to 
\begin{equation}
   \dot {\tilde \rho} = -i H_s \tilde \rho - \Gamma\tilde \rho \;.
\end{equation}
The same result may be obtained from the stochastic version of the HSR
model either through the cumulant expansion or through the application 
of the Shapiro-Loginov formula~\cite{shapiro78}.
Introducing the eigenstates of $H_s$
and using 
$\mu|0\rangle = \mu_0 \sum_n |n\rangle$,
the correlation function becomes,
\begin{eqnarray}
   \left \langle \mu(t) \mu(0) \right \rangle
   & = \sum_\kappa e^{-\Gamma t} e^{-i \omega_\kappa t} 
      \left\langle \kappa \left| \mu \right| 0 \right\rangle 
      \left \langle 0 \left | \mu \right| \kappa \right \rangle \nonumber \\
   & = e^{-\Gamma t} \sum_\kappa 
     \left |\mu_\kappa \right|^2
     e^{-i \omega_\kappa t} 
\end{eqnarray}
If $\Gamma=0$, then the Fourier transform of the dipole-dipole 
correlation function clearly leads back to~\eref{eq:disorder_spectra}.
At finite dephasing one obtains the result, 
\begin{equation}
   A(\omega) \propto \sum_\kappa 
   \left| \mu_\kappa \right|^2
   \frac{\Gamma}{\Gamma^2 + (\omega - \omega_\kappa)^2}
   \;,
   \label{eq:dephasing_spectra}
\end{equation}
which illustrates that the presence of dephasing simply broadens 
each line in the dephasing-free spectrum into a Lorentzian of width $\Gamma$.

At very large disorder, $\sigma/J\gg1$, the eigenstates are localized in the
site basis.
In this case, the dipole moments are all identical with $\mu_k = \mu_0$
which leads to a Gaussian spectrum in the zero dephasing limit.
As the dephasing increases the spectrum becomes a convolution of a Gaussian
with a Lorentzian --a Voigt profile.
In the alternative limit of very large dephasing, the absorption spectrum 
becomes independent of the disorder, 
as was also observed for the diffusion coefficients,
and the lineshape acquires a purely Lorentzian profile with a  
half maximum that scales linearly with the dephasing rate as seen
in figure~\ref{fig:fwhm}.

\section*{References}

\begin{thebibliography}{10}
\expandafter\ifx\csname url\endcsname\relax
  \def\url#1{{\tt #1}}\fi
\expandafter\ifx\csname urlprefix\endcsname\relax\def\urlprefix{URL }\fi
\providecommand{\eprint}[2][]{\url{#2}}

\bibitem{karl99}
Karl N, Kraft K, Marktanner J, M{\"u}nch M, Schatz F, Stehle R and Uhde H 1999
  {\em J. Vac. Sci. Technol. A\/} {\bf 17} 2318--2328

\bibitem{podzorov04}
Podzorov V, Menard E, Borissov A, Kiryukhin V, Rogers J~A and Gershenson M~E
  2004 {\em Phys. Rev. Lett.\/} {\bf 93} 086602

\bibitem{sakanoue10}
Sakanoue T and Sirringhaus H 2010 {\em Nature Materials\/} {\bf 9} 736--740

\bibitem{akselrod10}
Akselrod G~M, Tischler Y~R, Young E~R, Nocera D~G and Bulovic V 2010 {\em Phys.
  Rev. B\/} {\bf 82} 113106

\bibitem{dias09}
Dias F~B, Kamtekar K~T, Cazati T, Williams G, Bryce M~R and Monkman A~P 2009
  {\em {ChemPhysChem}\/} {\bf 10} 2096--2104

\bibitem{singh09}
Singh J, Bittner E~R, Beljonne D and Scholes G~D 2009 {\em J. Chem. Phys.\/}
  {\bf 131} 194905--194905--10

\bibitem{dykstra08}
Dykstra T~E, Hennebicq E, Beljonne D, Gierschner J, Claudio G, Bittner E~R,
  Knoester J and Scholes G~D 2008 {\em J. Phys. Chem. B\/} {\bf 113} 656--667

\bibitem{bednarz03}
Bednarz M, Malyshev V~A and Knoester J 2003 {\em Phys. Rev. Lett.\/} {\bf 91}
  217401

\bibitem{moix11}
Moix J, Wu J, Huo P, Coker D and Cao J 2011 {\em J. Phys. Chem. Lett.\/} {\bf
  2} 3045--3052

\bibitem{schofield95}
Schofield S~A and Wolynes P~G 1995 {\em J. Phys. Chem.\/} {\bf 99} 2753--2763

\bibitem{leitner96a}
Leitner D~M and Wolynes P~G 1996 {\em Phys. Rev. Lett.\/} {\bf 76} 216--219

\bibitem{leitner10}
Leitner D 2010 {\em New J. Phys.\/} {\bf 12} 085004

\bibitem{goj11}
Goj A and Bittner E~R 2011 {\em J. Chem. Phys.\/} {\bf 134} 205103--205103--11

\bibitem{cho05}
Cho M, Vaswani H~M, Brixner T, Stenger J and Fleming G~R 2005 {\em J. Phys.
  Chem. B\/} {\bf 109} 10542--10556

\bibitem{anderson58}
Anderson P~W 1958 {\em Phys. Rev.\/} {\bf 109} 1492

\bibitem{ishii73}
Ishii K 1973 {\em Prog. Theor. Phys. Supp.\/} {\bf 53} 77--138

\bibitem{phillips93}
Phillips P 1993 {\em Annu. Rev. Phys. Chem\/} {\bf 44} 115--144

\bibitem{kramer93}
Kramer B and {MacKinnon} A 1993 {\em Rep. Prog. Phys.\/} {\bf 56} 1469--1564

\bibitem{ingold04}
Ingold G, Wobst A, Aulbach C and H{\"a}nggi P 2004 What do phase space methods
  tell us about disordered quantum systems? {\em Anderson Localization and Its
  Ramifications\/} vol 630 ed Brandes T and Kettemann S (Springer Berlin /
  Heidelberg) pp 85--97

\bibitem{ping06}
Ping S 2006 {\em Introduction to Wave Scattering, Localization and Mesoscopic
  Phenomena\/} vol~88 (Springer)

\bibitem{kenkre82}
Kenkre V~M and Reineker P 1982 {\em Exciton Dynamics in Molecular Crystals and
  Aggregates\/} vol~94 (Springer)

\bibitem{madhukar77}
Madhukar A and Post W 1977 {\em Phys. Rev. Lett.\/} {\bf 39} 1424--1427

\bibitem{bulatov98}
Bulatov A, Kuklov A and Birman J~L 1998 {\em Chem. Phys. Lett.\/} {\bf 289}
  261--266

\bibitem{amir09}
Amir A, Lahini Y and Perets H~B 2009 {\em Phys. Rev. E\/} {\bf 79} 050105

\bibitem{izrailev97}
Izrailev F~M, Kottos T, Politi A and Tsironis G~P 1997 {\em Phys. Rev. E\/}
  {\bf 55} 4951--4963

\bibitem{thouless81}
Thouless D~J and Kirkpatrick S 1981 {\em Journal of Physics C: Solid State
  Physics\/} {\bf 14} 235--245

\bibitem{mott69}
Mott N~F 1969 {\em Philosophical Magazine\/} {\bf 19} 835--852

\bibitem{logan87}
Logan D~E and Wolynes P~G 1987 {\em Phys. Rev. B\/} {\bf 36} 4135--4147

\bibitem{mukamel89}
Mukamel S 1989 {\em Phys. Rev. B\/} {\bf 40} 9945--9947

\bibitem{loring88}
Loring R~F, Franchi D~S and Mukamel S 1988 {\em Phys. Rev. B\/} {\bf 37}
  1874--1883

\bibitem{evensky90}
Evensky D~A, Scalettar R~T and Wolynes P~G 1990 {\em J. Phys. Chem\/} {\bf 94}
  1149--1154

\bibitem{yamada99}
Yamada H and Ikeda K~S 1999 {\em Phys. Rev. E\/} {\bf 59} 5214--5230

\bibitem{cao09}
Cao J and Silbey R~J 2009 {\em J. Phys. Chem. A\/} {\bf 113} 13825--13838

\bibitem{hoyer10}
Hoyer S, Sarovar M and Whaley K~B 2010 {\em New J. Phys.\/} {\bf 12} 065041

\bibitem{miller60}
Miller A and Abrahams E 1960 {\em Phys. Rev.\/} {\bf 120} 745--755

\bibitem{lloyd11}
Lloyd S, Mohseni M, Shabani A and Rabitz H 2011 {\em {arXiv:1111.4982}\/}

\bibitem{wu12b}
Wu J, Cao J and Silbey R~J 2013 {\em Phys. Rev. Lett., in press\/}

\bibitem{wu12}
Wu J, Liu F, Ma J, Silbey R~J and Cao J 2012 {\em J. Chem. Phys.\/} {\bf 137}
  174111--174111--12

\bibitem{ciuchi11}
Ciuchi S, Fratini S and Mayou D 2011 {\em Phys. Rev. B\/} {\bf 83} 081202

\bibitem{lee85}
Lee P~A and Ramakrishnan T~V 1985 {\em Reviews of Modern Physics\/} {\bf 57}
  287--337

\bibitem{mukamel99}
Mukamel S 1999 {\em Principles of Nonlinear Optical Spectroscopy\/} (Oxford
  University Press, {USA})

\bibitem{witkoskie02}
Witkoskie J~B, Yang S and Cao J 2002 {\em Phys. Rev. E\/} {\bf 66} 051111

\bibitem{jayannavar88}
Jayannavar A~M and Kumar N 1988 {\em Phys. Rev. B\/} {\bf 37} 573--576

\bibitem{moix12}
Moix J~M, Zhao Y and Cao J 2012 {\em Phys. Rev. B\/} {\bf 85} 115412

\bibitem{trambly_de_laissardiere06}
de~Laissardière G~T, Julien J and Mayou D 2006 {\em Phys. Rev. Lett.\/} {\bf
  97} 026601

\bibitem{klafter78}
Klafter J and Jortner J 1978 {\em J. Chem. Phys.\/} {\bf 68} 1513--1522

\bibitem{fidder91a}
Fidder H, Knoester J and Wiersma D~A 1991 {\em J. Chem. Phys.\/} {\bf 95}
  7880--7890

\bibitem{wu10}
Wu J, Liu F, Shen Y, Cao J and Silbey R~J 2010 {\em New J. Phys.\/} {\bf 12}
  105012

\bibitem{valleau12}
Valleau S, Saikin S~K, Yung M and Guzik A~A 2012 {\em J. Chem. Phys.\/} {\bf
  137} 034109--034109--13

\bibitem{zhong13b}
Zhong X, Zhao Y and Cao J unpublished results

\bibitem{lee06}
Lee K, Cho S, Park S~H, Heeger A~J, Lee C and Lee S 2006 {\em Nature\/} {\bf
  441} 65--68

\bibitem{escalante10}
Escalante M, Lenferink A, Zhao Y, Tas N, Huskens J, Hunter C~N, Subramaniam V
  and Otto C 2010 {\em Nano Letters\/} {\bf 10} 1450--1457

\bibitem{zhong13}
Zhong X and Zhao Y 2013 {\em J. Chem. Phys.\/} {\bf 138} 014111--014111--9

\bibitem{rodriguez03}
Rodr{\'i}guez A, Malyshev V~A, Sierra G, {Martín-Delgado} M~A,
  {Rodríguez-Laguna} J and {Domínguez-Adame} F 2003 {\em Phys. Rev. Lett.\/}
  {\bf 90} 027404

\bibitem{mukamel85}
Mukamel S 1985 {\em J. Chem. Phys.\/} {\bf 82} 5398--5408

\bibitem{shapiro78}
Shapiro V~E and Loginov V~M 1978 {\em Physica A\/} {\bf 91} 563--574

\end{thebibliography}
\providecommand{\newblock}{}

\end{document}